\documentclass[aps,prd,11pt,superscriptaddress,floatfix,nofootinbib,notitlepage]{revtex4-1} 
\usepackage{colortbl}
\usepackage[T1]{fontenc} 
\usepackage{lmodern}
\usepackage{hyperref}
\usepackage{graphicx}
\usepackage{bm}
\usepackage{amsmath}
\usepackage{amssymb}
\usepackage{dcolumn}
\usepackage{multirow}
\usepackage{color}
\usepackage[dvipsnames]{xcolor}
\usepackage[papersize={8.5in,11in},top=2cm, bottom=1.5cm, left=1.5cm, right=1.5cm]{geometry}

\begin{document} 
\title{On The Feasibility Of Using Neutrino Intensity Interferometry To Measure Proto-Neutron Star Radii}

\author{Warren P. \surname{Wright}}
\email{wpwright@ncsu.edu}
\affiliation{Department of Physics,  North Carolina State University, Raleigh, North Carolina 27695, USA}

\author{James P. \surname{Kneller}}
\email{jpknelle@ncsu.edu}
\affiliation{Department of Physics,  North Carolina State University, Raleigh, North Carolina 27695, USA}

\date{\today}

\begin{abstract}

It has recently been demonstrated analytically that the two-point correlation function for pairs of neutrinos may contain information about the size of the proto-neutron star formed in a Galactic core-collapse supernova. The information about the size of the source emerges via the neutrino equivalent of intensity interferometry originally used by Hanbury-Brown and Twiss with photons to measure the radii of stars. However the analytic demonstration of neutrino intensity interferometry with supernova neutrinos made a number of approximations: that the two neutrinos had equal energies, the neutrinos were emitted at simultaneous times from two points and were detected simultaneously at two detection points that formed a plane with the emission points. These approximations need to be relaxed in order to better determine the feasibility of neutrino intensity interferometry for supernovae neutrinos in a more realistic scenario. In this paper we further investigate the feasibility of intensity interferometry for supernova neutrinos by relaxing all the approximations made in the earlier study. We find that, while relaxing any one assumption reduces the correlation signal, the relaxation of the assumption of equal times of detection is by far the largest detrimental factor. For neutrino energies of order $\sim 15\;{\rm MeV}$ and a supernova distance of $L = 10\;{\rm kpc}$, we show that in order to observe the interference pattern in the two-point correlation function of the neutrino pairs, the timing resolution of a detector needs to be on the order of $\lesssim 10^{-21}\;{\rm s}$ if the initial neutrino wave packet has a size of $\sigma_x \sim 10^{-11}\;{\rm cm}$. 
\end{abstract}

\maketitle

\section{Introduction \label{Introduction}}

The neutrino signal from a Galactic core-collapse supernova is expected to rich with information about the sequence of events that occur during the explosion and with information about the properties of the neutrino. The information is imprinted via the time, energy and flavor composition of the signal. For recent reviews of what we may learn from the neutrino signal from the next Galactic supernova we refer the reader to the reviews by Scholberg \cite{2012ARNPS..62...81S}, Mirrizzi \emph{et al.} \cite{2016NCimR..39....1M}, and Horiuchi \& Kneller \cite{2018JPhG...45d3002H}. In addition, it was recently shown by Wright \& Kneller \cite{WrightNII} - hereafter W\&K - that there might also be information about the supernova and the neutrino in another channel: the separation in space between simultaneously detected events. The origin of this effect is simply the interference between the two possible pairs of paths from two emission points on the neutrinosphere to the two detection points. This is the same interference effect used originally by Hanbury-Brown and Twiss \cite{1956Natur.177...27B} with photons to measure the radii of stars \cite{1956Natur.178.1046H} and is known as HBT or Intensity Interferometry. This technique has since been used to measure the emission region of many other systems, see Baym \cite{1998AcPPB..29.1839B} for examples, and it has also been previously suggested as a method for determining the Majorana or Dirac nature of the neutrino \cite{2006PhRvL..96l1802G}. 

In a simplified calculation, W\&K showed how intensity interferometry using supernova neutrinos could be used to determine the size of the source - i.e. the neutrinosphere - for a supernova at a distance of $10\;{\rm kpc}$ using neutrinos with an energy of order $E \sim 10\;{\rm MeV}$ in detectors with dimensions of order tens to hundreds of meters. They found that as long as the initial neutrino wave packet was not smaller than $\sim 10^{-12}\;{\rm cm}$ and the neutrino mass was not less than $\sim 10^{-8}\;{\rm eV}$, spatial variation of the two-particle correlation function was visible on the scale of typical neutrino detector dimensions and thus it seemed possible to measure the neutrinosphere radius given sufficient statistics and detector time resolution. However in order to draw that conclusion, W\&K made a number of approximations in order to make their analysis tractable. They assumed: the two neutrinos had equal energies, that the neutrinos were emitted at simultaneous times from just two points, and that they were detected simultaneously at two detection points that formed a plane with the two points of emission. These approximations need to be relaxed in order to determine whether W\&K's optimism that the technique could yield useful information is justified within a more realistic scenario. 

The goal of this paper is to further explore the phenomenon of neutrino intensity interferometry and its application to supernova neutrinos by relaxing the assumptions that went in to the W\&K analysis. To investigate the larger parameter space we construct a sample of event pairs using a Markov Chain Monte Carlo algorithm and compute the statistics of the sample as a function of eleven parameters that describe the two-particle wave packet. Our paper is structured as follows: in section \S\ref{sec:Analytical_Analysis} we derive the expression for the two-particle correlation function and then describe the Markov-Chain Monte Carlo algorithm used to determine the expected pattern of simultaneous events as a function of the eleven parameters that enter into the expression. In section \S\ref{sec:results} we present the numerical results of the sample we generated and then explain analytically in section \S\ref{sec:analytical} the most important finding. We discuss the results and conclude in section \S\ref{sec:Conclusion}. 


\section{Expanded Analysis \label{sec:Analytical_Analysis}}
We begin with the wave packet formulation of the single-particle wave function ($\psi_{ij}$) given in Equation 2 of W\&K for a neutrino with mass $m_\nu$ created with quantum limited position and momentum uncertainty, $\sigma_x$ and $\sigma_p$ respectively (i.e. $2\,\sigma_x\,\sigma_p=1$, in natural units), emitted from spacetime point $t_{ri},\vec{r}_{i}$ with energy $E_i$ and detected at spacetime point $t_{dj},\vec{d}_{j}$:
\begin{equation}
	\psi_{ij}\equiv\psi_{\vec{p}_{ij}}\left(\vec{x}_{ij},t_{ij}\right)=
		\frac{\left(2\pi\right)^{-3/4}}{\sigma_
    		{\perp ij}\sqrt{\sigma_{\parallel ij}}}
        \,\text{Exp}\left(
        	\bm{i}\left(p_{ij}\cdot x_{ij}\right)
			-\frac{\vec{b}_{ij}^2}{4\sigma_x\sigma_{\perp ij}}
			-\frac{\bm{i}t_{ij}\left(\vec{b}_{ij}\cdot\vec{p}_{ij}\right)^2}
			{8E_i^3\sigma_x^2\,\sigma_{\perp ij}\,\sigma_{\parallel ij}}
        \right).
    \label{Eqn:WF}
\end{equation}
In this equation $\vec{p}_{ij}$ is the central momentum of the neutrino wave packet, $\vec{x}_{ij} = \vec{d}_j - \vec{r}_i$ is the displacement and $t_{ij} = t_{dj} - t_{ri}$ is the time elapsed from when the center of the wave packet was at $\vec{r}_{i}$ to when the neutrino was detected at $\vec{d}_{j}$. Additionally we define the quantities $\sigma_{\perp ij}=\sigma_x+\bm{i}\,t_{ij}\,\sigma_p/E_i$ which is the lateral spread of the wave packet, $\sigma_{\parallel ij}=\sigma_x+\bm{i}\,t_{ij}\,\sigma_p/(E_i \gamma_i^2)$ which is the longitudinal spread of the wave packet, $\vec{b}_{ij}=\vec{x}_{ij}-t_{ij}\,\vec{p}_{ij}/E_i$ which is the spatial offset of the detection point from the path of the wave packet centroid, and the Lorentz factor is $\gamma_i=E_i/m_{\nu}$. Note that in what follows we shall ignore flavor oscillations in the supernova mantle - collective flavor oscillations are suppressed during the early phases of the supernova due to the large matter density \cite{2011PhRvL.107o1101C,2012PhRvL.108f1101S} - and do not take into account the misalignment between the neutrino mass and flavor states in the detection process. Such details will not greatly affect our results.

\begin{figure}[ht]
	\includegraphics[width=0.5\linewidth]{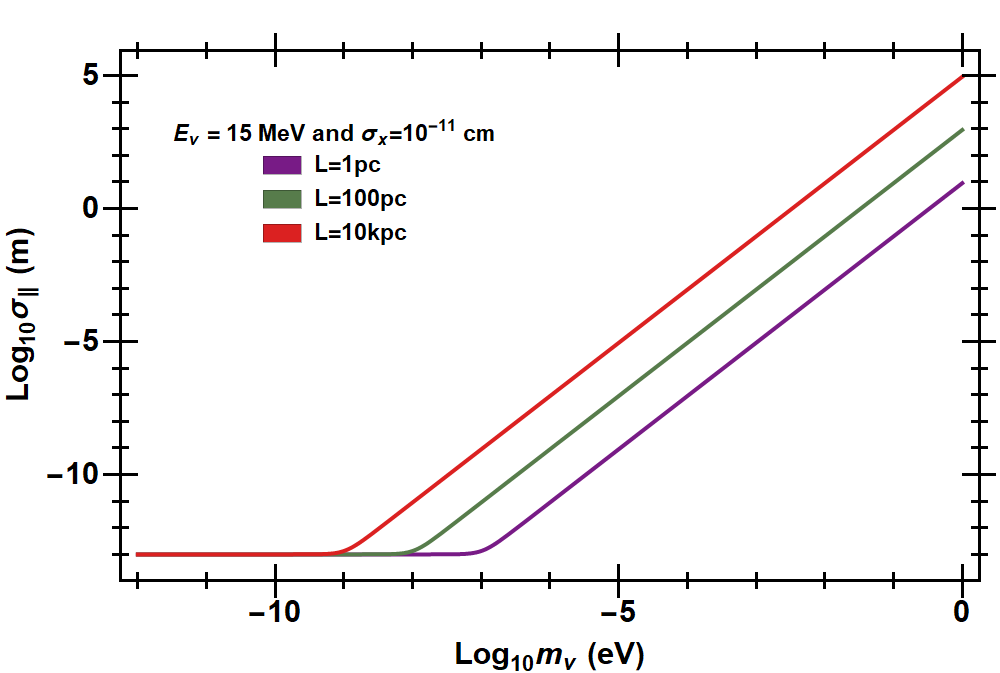}
	\caption{Growth of the longitudinal thickness of the wave packet.}
	\label{fig:WPgrowthParallal}
\end{figure}

The increase in the size of the neutrino wave packet over an astronomical distance can be significant. The longitudinal size of the neutrino wave packet for a neutrino energy of $E=15\;{\rm MeV}$ and an initial wave packet size of $\sigma_x=10^{-11}\;{\rm cm}$ as a function of a the neutrino mass and for various supernova distances is shown in Fig. \ref{fig:WPgrowthParallal}. Notice how the longitudinal spread of the wave packet decreases as the neutrino mass decreases for a given supernova distance but that it has a floor of $\sigma_x$. For supernovae at a distance in the range of 1 pc to 10 kpc and a neutrino energy around 10 MeV, the longitudinal spread of the neutrino wavepacket at Earth is much greater than $\sigma_x$ if the neutrino mass is greater than $10^{-9}\;{\rm eV}$. In the limit where $\sigma_{\parallel} \gg \sigma_x$ and $\sigma_{\perp} \gg \sigma_x$, the scaling of $\sigma_{\parallel}$ and $\sigma_{\perp}$ follows
\begin{eqnarray}
\sigma_{\perp} & \approx & \left(66\text{ pc}\right)\left(\frac{L}{1\; \text{kpc}}\right)\,\left(\frac{15\;{\rm MeV}}{E} \right)\,\left(\frac{10^{-11}\;{\rm cm}}{\sigma_x} \right) \\
\sigma_{\parallel} & \approx & \left(9\text{ km}\right)\left(\frac{L}{1\; \text{kpc}}\right)\,\left(\frac{m_\nu}{1\; \text{eV}}\right)^2\,\left(\frac{15\;{\rm MeV}}{E} \right)^3\,\left(\frac{10^{-11}\;{\rm cm}}{\sigma_x} \right)
\end{eqnarray}
Counter-intuitively, the larger the spatial size of the wave packet at the source, the smaller it is in the limit where $\sigma_{\parallel} \gg \sigma_x$ and $\sigma_{\perp} \gg \sigma_x$. The enormous growth in the size of the neutrino wave packet is why their overlap, and consequent interference, must be considered. A rough estimate of the number of overlapping wave packets in a detector can be made. Two neutrinos detected simultaneously with a separation along the line of sight to the supernova of $\sigma_{\parallel}$ or less will have had overlapping wave packets. The number $N_{2\nu}$ of overlapping wave packets per unit area is thus $N_{2\nu} \sim F \sigma_{\parallel} / c$ where $F$ is the neutrino flux at Earth and $c$ the speed of light. For a supernova at $L = 10\;{\rm kpc}$ emitting $10^{58}$ neutrinos over a 10 second period, the flux $F$ is of order $F \sim 10^{15}\;{\rm /m^2/s}$. Thus the estimate for the number of overlapping wave packets is $N_{2\nu} \sim 10^{12}\;{\rm /m^2}$ for a neutrino mass of $m_{\nu} = 1\;{\rm eV}$. 

Intensity Interferometry is the effect which occurs when there are alternative multi-particle wavefunctions connecting the points of emission to the points of detection. For two neutrinos emitted from points $\vec{r}_1$ and $\vec{r}_2$ and detected at $\vec{d}_1$ and $\vec{d}_2$, the two-particle wavefunction, given in W\&K, is 
\begin{align}
	\phi_{\vec{p}_1,\vec{p}_2}\left(\vec{r}_1,\vec{d}_1,\vec{r}_2,\vec{d}_2\right) = 
		\frac{1}{\sqrt{2}}\left(\psi_{11}\,\psi_{22} -\psi_{12}\,\psi_{21}\right).
    \label{Eqn:TwoParticleWF}
\end{align}
where the single particle wave-functions $\psi_{ij}$ are given in Eq. \ref{Eqn:WF}. The two particle probability density is
\begin{align}
    \vert\phi_{\vec{p}_1,\vec{p}_2}\vert^2=&\frac{1}{2}\left(
        \vert\psi_{11}\vert^2\,\vert\psi_{22}\,\vert^2
        +\vert\psi_{12}\vert^2\,\vert\psi_{21}\,\vert^2
    \right)-\frac{1}{2}\left(
        \psi_{11}^* \, \psi_{22}^*\,\psi_{12} \, \psi_{21}
        +\psi_{12}^* \, \psi_{21}^*\,\psi_{11}\,\psi_{22}
    \right).
    \label{Eqn:TwoParticlePDensity}
\end{align}
and the two point correlation function (2PCF), $C_2(d)$, the quantity one would hope to measure in an experiment, is given by 
\begin{equation}
C_2 = \frac{2\, \vert\phi\vert^2}{\vert\psi_{11}\vert^2\,\vert\psi_{22}\vert^2 +\vert\psi_{12}\vert^2\,\vert\psi_{21}\vert^2} 
= 1 - \frac{ \left( \psi_{11}^*\,\psi_{22}^*\,\psi_{12}\,\psi_{21} - \psi_{12}^*\,\psi_{21}^*\,\psi_{11}\,\psi_{22} \right) }
        { \vert\psi_{11}\vert^2\,\vert\psi_{22}\vert^2 +\vert\psi_{12}\vert^2\,\vert\psi_{21}\vert^2 }. 
\end{equation}        
The question becomes whether the 2PCF is observable. To answer this question W\&K made a number of approximations in order to determine the observability of the interference pattern in the 2PCF with a reduced set of parameters. They found that if they set both emission times and both detection times to be equal i.e. $t_{r1}=t_{r2}$ and $t_{d1}=t_{d2}$, assumed equal energies for the neutrinos, and confined the geometry to a plane, the 2PCF has a spatial variation which could be observed on the scale of tens of meters for $15\;{\rm MeV}$ neutrinos emitted from two points separated by tens of kilometers from a source at a distance of $10\;{\rm kpc}$. Our goal is to relax these assumptions and allow for two different neutrino energies, non-coincident times of emission from a hemispherical source, non-coincident times of detection, and a non-planar geometry. 

\subsection{Ensemble Generation} \label{sec:MCMC_Description}
In order to determine whether the 2PCF is sensitive to the parameters that enter into the two particle wave function, we generate an ensemble of event pairs via a Markov-Chain Monte-Carlo (MCMC) method based upon the Metropolis-Hastings algorithm. After generating the sample, we can examine the distribution of the events with respect to each of parameters separately, but also we can study the distribution of the event pairs as a function of pairs of parameters. The pair we are most interested in is the sensitivity of the 2PCF as a function of $R$, the source radius, and $d$, the event separation. As an expectation, fermion statistics tells us that, regardless of $R$, there should be no events at $d=0$ if the detection times and energies are equal. 

We set the distance to the supernova, $L$, the initial wave packet size, $\sigma_{x}$, and the neutrino mass $m_{\nu}$ to be fixed which leaves eleven independent parameters needed to define the two-particle wave packet. They are: $R$ which is the radius of the neutrinosphere, $\theta_1, \theta_2, \phi_1$ and $\phi_2$ which define the two initial positions on the neutrinosphere that emitted the two detected neutrinos; $\Delta E_1$ and $\Delta E_2$ which define the neutrinos energies via $E_i=E_\text{mid}+\Delta E_i$ where $E_\text{mid}=15$ MeV for our analysis; and lastly; $d$ which is the separation of the events in the detector. We align the center of the proto-neutron star with the origin and place the center of the detector along the z-axis. The detected events are taken to occur along a line parallel to the x axis direction - the direction of the event pairs may be fixed this way due to the rotational symmetry of the problem which  means the relevant quantity is the relative angle between the axis of the detected events and the axis of the emission points, not the orientation of each separately. Finally, there are four times to consider: the two times of emission at the source, $t_{r1}$ and $t_{r2}$, and two times of detection, $t_{d1}$ and $t_{d2}$, but, without loss of generality we can set one of the detection times to be the propagation time of the centroid of the wave packet between the source and the detector and then label the second detection time by the lapse $\Delta t_{d}$. Thus the detection times are $t_{d1}=L/v$ and $t_{d2}=L/v+\Delta t_{d}$ where $v$ is the neutrino velocity for $E_\text{mid}$. 

Each of the 11 parameters is examined over a bounded interval. The natural range of the four angles are $0<\theta_i<2\pi$ and $0<\phi_i<\pi/2$ and we adopt uniform distributions for these angles i.e. the neutrinos have a half-isotropic distribution of emission angles at the neutrinosphere and we do not consider limb darkening. We adopt the range $0<d<300$ m also with a uniform distribution in order to cover the dimensions of current and future neutrino detectors. Given model expectations for the neutrinosphere radius we set $0<R<80$ km. Finally, for $\Delta E_1$, $\Delta E_2$, $t_{r1}$, $t_{r2}$ and $\Delta t_{d}$ the bounds on these quantities are fixed for each run of the MCMC and we shall consider many different ranges. The bounds will be listed as the results are presented. 

At each iteration of the algorithm, the new values of the eleven variable parameters are drawn from a truncated Normal Distribution with a mean given by the previous value of the parameter and the standard deviation is the width of the parameter's interval divided by a scale factor. Special care has been taken to ensure that the probability of accepting a new proposal is correctly modified by an acceptance factor if any of the parameters have values near their limits. The two-particle wavefunction at the new location in the parameter space is computed and if it is more probable than the last location, the new location in the parameter space is accepted and added to the chain. If it is less probable than the last location, the ratio of the probability at the new location relative to the previous location is tested against a uniform random number. If the random number is less than the ratio, the new location is also added to the chain but if the random number is larger than the ratio of probabilities, the new location is rejected and the algorithm retains the previous location and attempts an alternative trial location in the next iteration.

For the Metropolis-Hastings algorithm, an initial point needs to be assigned. Our approach is to choose three evenly spaced values in each parameter's interval to define our set of initial values. Thus we have $3^{11}=177,147$ initial values and each of these is used to initialize the first link in the Markov chain (also known as a ``walker") of the algorithm. All of the chains we make contain 1000 links. Initially the location of the chain links in the parameter space will be biased toward the initial point and so the common practice is to "burn" these biased iterations accomplished, in our case, by discarding the first 250 links. Further details about the dependence of the algorithm's convergence on the scale-factor, the length of the chains, and the burn count is provided in Appendix \ref{appendix:MCMC_CandC}.

\section{Results}\label{sec:results}

\begin{figure}[b]
	\includegraphics[width=0.5\linewidth]{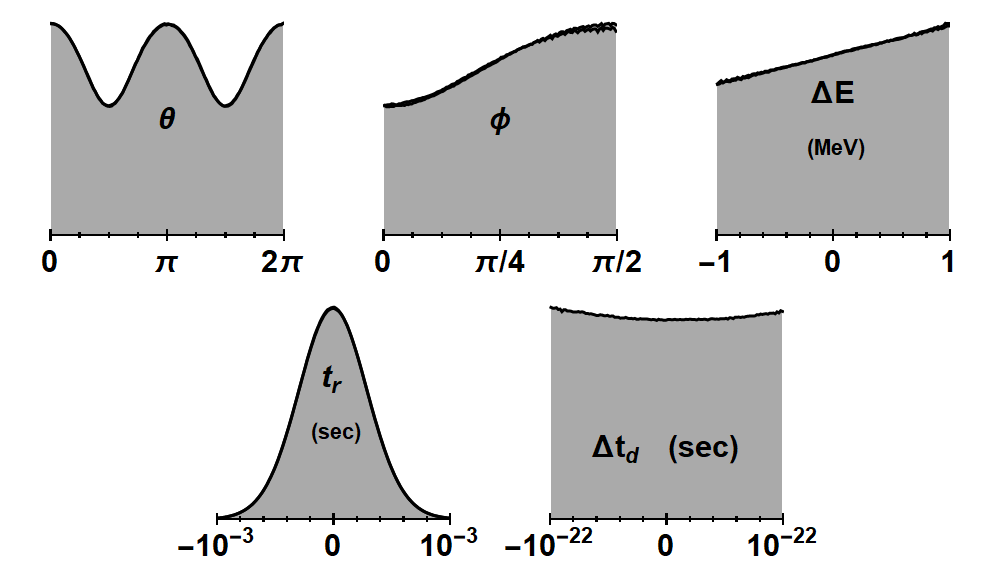}
	\caption{Histograms of the number of events as a function of the variable $\theta$, $\phi$, $\Delta E_i$, $t_{ri}$ and $\Delta t_{d}$ when $\Delta E_i$ is constrained to be $|\Delta E_i|\leq 1\;{\rm MeV}$, $|t_{ri}|\leq 10^{-3}\;{\rm s}$ and $|\Delta t_{d}| \leq 10^{-22}\;{\rm s}$.}
	\label{fig:MCMC_Histograms_td-22_1D}
\end{figure}

\subsection{Uniform Event Distributions}
\begin{figure}[t]
	\includegraphics[width=0.5\linewidth]{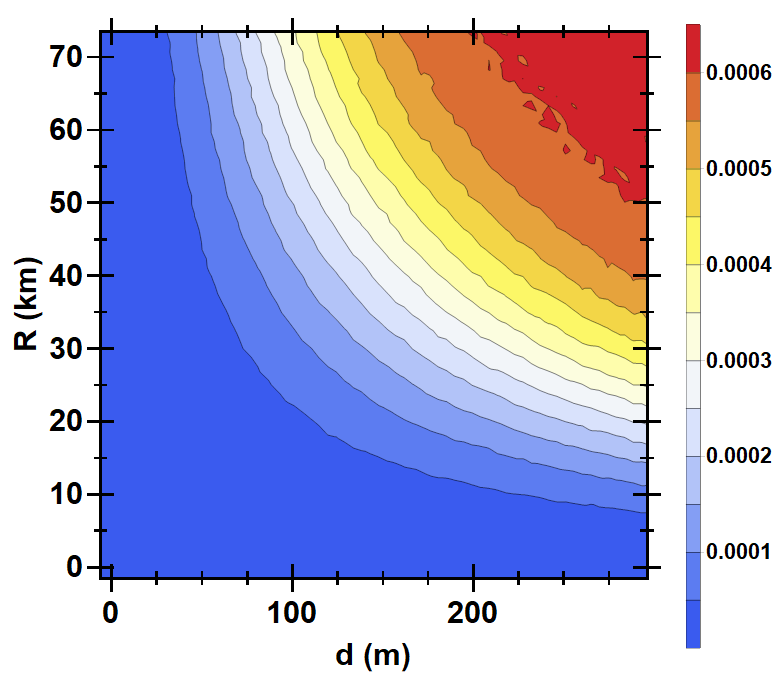}
	\caption{Normalized event counts as a function of the source radius $R$ and the separation in the detector $d$ for the case when $\Delta E_i$ is constrained to be $|\Delta E_i|\leq 1\;{\rm MeV}$ $|t_{ri}|\leq 10^{-3}\;{\rm s}$ and $|\Delta t_{d}| \leq 10^{-22}\;{\rm s}$.}
	\label{fig:MCMC_Histograms_td-22_2D}
\end{figure}

We first consider a setup where the non-variable parameters take the following values: $L=10$ kpc, $\sigma_x=10^{-11}$ cm and $m_\nu=1$ eV and the variable parameters are bound to the following intervals $|\Delta E_i|<1$ MeV, $|t_{ri}|<10^{-3}$ s. The distribution for the neutrinosphere radius is taken to be uniform with all values equally likely. Lastly, we restrict ourselves to the rather extreme case where and $|\Delta t_{d}|<10^{-22}$ s. With such a small permitted difference between the detection times we can isolate the effects of allowing neutrino emission from a two random points on a hemisphere and with similar, but not identical energies. As described above, we initialize $3^{11}$ chains, iterate each 1000 times using a scale factor of 20 and lastly, we remove the first 250 links of each chain. The results of this computation are shown in Fig. \ref{fig:MCMC_Histograms_td-22_1D} and Fig. \ref{fig:MCMC_Histograms_td-22_2D}. For each histogram, the y-axis is the number of counts in each of 128 uniform bins within the range of each parameter and displays values from zero counts up to the maximum number ($\sim10^6$). The contour plot of the 2D histogram shows the normalized event pair sample in $64 \times 64$ bins. Each of the histograms displayed in Fig. \ref{fig:MCMC_Histograms_td-22_1D} and Fig. \ref{fig:MCMC_Histograms_td-22_2D} will be discussed in turn.

\begin{figure}[b]
	\includegraphics[width=0.5\linewidth]{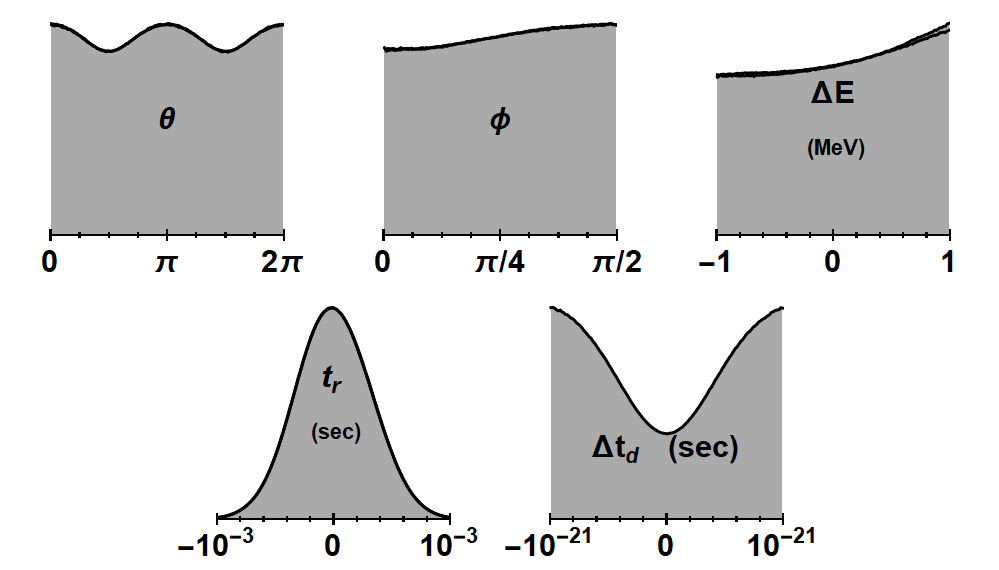}
	\caption{The same as figure (\ref{fig:MCMC_Histograms_td-22_1D}) but for the case $|\Delta E_i|\leq 1\;{\rm MeV}$ $|t_{ri}|\leq 10^{-3}\;{\rm s}$ and $|\Delta t_{d}| \leq 10^{-21}\;{\rm s}$
    }
	\label{fig:MCMC_Histograms_td-21_1D}
\end{figure}

\begin{figure}[h!!!]
	\includegraphics[width=0.5\linewidth]{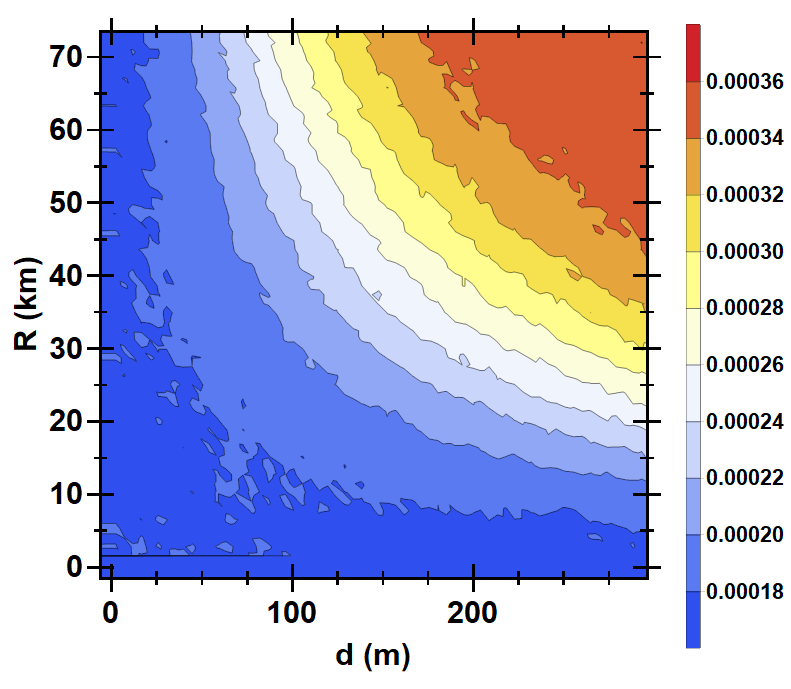}
	\caption{Normalized event counts as a function of the source radius $R$ and the separation in the detector $d$ for the case $|\Delta E_i|\leq 1\;{\rm MeV}$ $|t_{ri}|\leq 10^{-3}\;{\rm s}$ and $|\Delta t_{d}| \leq 10^{-21}\;{\rm s}$
    }
	\label{fig:MCMC_Histograms_td-21_2D}
\end{figure}

\begin{figure}[b]
	\includegraphics[width=0.5\linewidth]{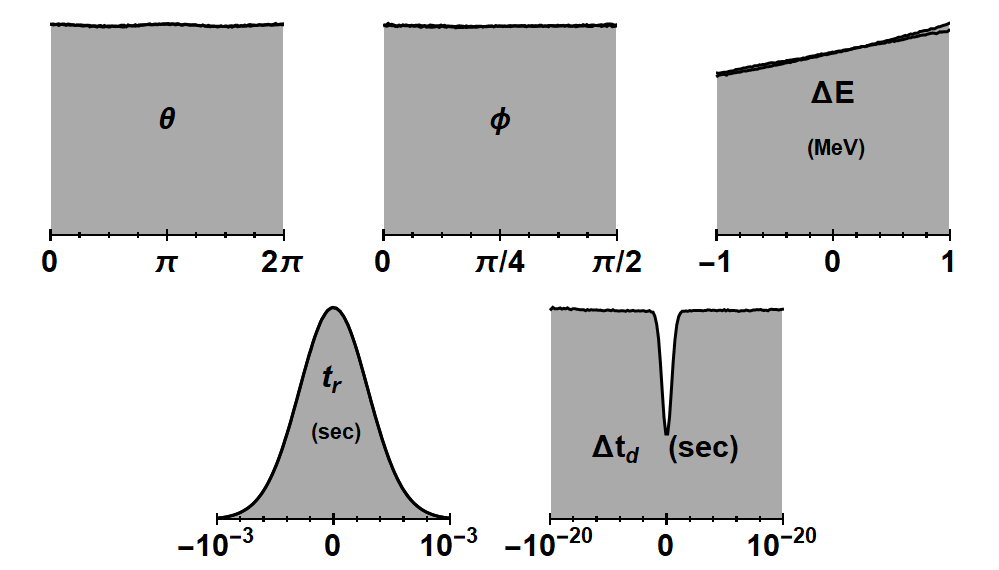}
	\caption{The same as figure (\ref{fig:MCMC_Histograms_td-22_1D}) but for the case $|\Delta E_i|\leq 1\;{\rm MeV}$ $|t_{ri}|\leq 10^{-3}\;{\rm s}$ and $|\Delta t_{d}| \leq 10^{-20}\;{\rm s}$
    }
	\label{fig:MCMC_Histograms_td-20_1D}
\end{figure}

\begin{figure}[h!!!]
	\includegraphics[width=0.5\linewidth]{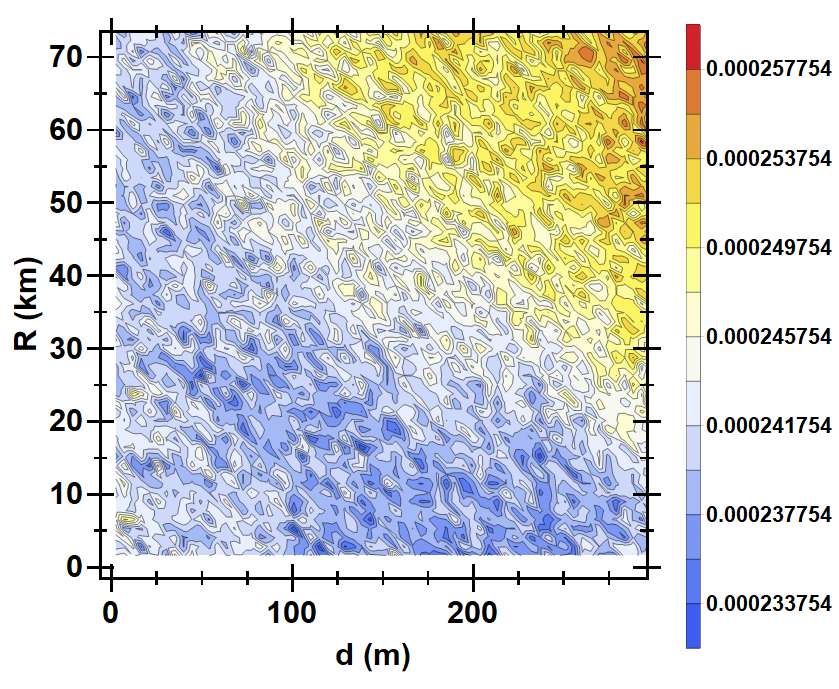}
	\caption{Normalized event counts as a function of the source radius $R$ and the separation in the detector $d$ for the case $|\Delta E_i|\leq 1\;{\rm MeV}$ $|t_{ri}|\leq 10^{-3}\;{\rm s}$ and $|\Delta t_{d}| \leq 10^{-20}\;{\rm s}$
    }
	\label{fig:MCMC_Histograms_td-20_2D}
\end{figure}

The $\theta$ histogram in Fig. \ref{fig:MCMC_Histograms_td-22_1D} shows the combined distribution of $\theta_1$ and $\theta_2$. The peaks in the distribution of this parameter occur at the angles where the axis of emission of the two neutrinos lies in the same plane as the detection. These peaks indicate the two-particle wave packet probability density is largest when the points of emission and detection form a plane. Conversely, the troughs indicate the two particle probability density is smaller when the neutrinos are emitted in a plane perpendicular to the detection plane. If we allow for the alignment of the detection axis to be arbitrary these peaks and troughs would disappear for a source emitting uniformly. However, simulations of core-collapse in multi-dimensions find the neutrino emission during the first second post-bounce can be anisotropic either due a large scale asymmetry such as the Standing Accretion Shock Instability \cite{2003ApJ...584..971B,2013ApJ...770...66H,2015MNRAS.452.2071F} or Lepton-Emission Self-sustained Asymmetry (LESA) \cite{2014ApJ...792...96T,Tamborra:2014hga,2017ApJ...839..132T}, or small scale anisotropy due to neutrino emission `hotspots' at the base of `downflows,' \cite{2015ApJS..216....5S,2016ApJ...818..123B}.

The $\phi$ histogram shows the two overlapping histograms of $\phi_1$ and $\phi_2$. Interestingly this distribution indicates that dual neutrino detection is more probable for neutrinos emitted from near the edge as opposed to the center of the emitting hemisphere. Similarly, the $\Delta E$ histogram shows the two histograms for the $\Delta E_1$ and $\Delta E_1$ variables. The increase in probability for higher energies is not due to the energy dependence of neutrino cross sections - an effect which would also skew event pairs to those with higher energies - and will be briefly commented on in Sec. \ref{sec:Analytical_Analysis}. The $\Delta t_d$ histogram is qualitatively flat but the $t_r$ histogram, which is the combined distribution of $t_{r1}$ and $t_{r2}$, and has clear structure. The Gaussian-like shape of the $t_{ri}$ histogram indicates the algorithm fully covered the time range that could lead to overlapping neutrino wave packets in a detector for our given setup. The radius of the meutrinosphere was restricted to $R < 80\;{\rm km}$ which corresponds to a maximum light travel time of $2.7\times 10^{-4}\;{\rm s}$. Thus our results indicate the time window for emission is a factor of a few times the light crossing time of the source i.e. $\Delta t_r = |t_{r1} - t_{r2}| \lesssim few\, R/c$.

Lastly, the 2D histogram (Fig. \ref{fig:MCMC_Histograms_td-22_2D}) of the $R$ and $d$ variables shows the hoped for correlation signal. The fact that the contours vary both in $R$ and $d$ indicates that the pattern of spacial separation of two-particle detection events is large and furthermore, it changes with source size. The goal of any static-source, intensity-interferometry experiment is to measure the events-vs-separation distribution and, by fitting it to a predictive model (such as the 2D histogram in Fig. \ref{fig:MCMC_Histograms_td-22_2D}), determine the source size. Our results show that allowing the neutrinos to be emitted from a hemisphere at different times and with unequal energies does not eradicate the information about the source size in the 2PCF seen in the simpler analysis by W\&K. 

Let us now increase the size of the detection time window $\Delta t_d$.
Figures \ref{fig:MCMC_Histograms_td-21_1D} to \ref{fig:MCMC_Histograms_td-20_2D} show the same quantities as displayed in Figs. \ref{fig:MCMC_Histograms_td-22_1D} and \ref{fig:MCMC_Histograms_td-22_2D}. The difference is that for Fig. \ref{fig:MCMC_Histograms_td-21_1D} and \ref{fig:MCMC_Histograms_td-21_2D}, the detection time window has been broadened to $|\Delta t_{d}|<10^{-21}$ s. While the distribution of the event pairs with respect to $t_r$ and $\Delta E$ are largely unchanged, the peaks and troughs in the distribution for the angles $\theta_1$ and $\theta_2$ are now much smaller, the preference for event pairs which are emitted towards the edge of the disk is also less pronounced, and the distribution of detector time separation now has a clear minimum at zero. More importantly, the 2D contour plot for the distribution of events with the radius $R$ and event separation $d$ of the sample is now more uniform across the plane. The reduction of the variance in this joint distribution indicates the broader time detection window makes the hoped for signal weaker and more difficult to detect. In order to determine a neutrinosphere radius we would need many more pairs of events than for the previous case shown in Fig. \ref{fig:MCMC_Histograms_td-22_2D}. 

Finally, in Figs. \ref{fig:MCMC_Histograms_td-20_1D} and \ref{fig:MCMC_Histograms_td-20_2D}, the detection time window has been broadened to $|\Delta t_{d}|<10^{-20}$ s. Again the distribution of the event pairs with $t_{r1}$ and $t_{r2}$ still prefers the case when the neutrinos are emitted simultaneously and the distribution with energy is also the same as the previous results with a preference for energies slightly higher than $E_{mid}$ rather than below. But for the other variables the distribution of the angles $\theta_1$/$\theta_2$ and $\phi_1$/$\phi_2$ are completely uniform, and the minimum in $\Delta t_{d}$ at $\Delta t_{d}=0$ is seen to be a feature that occurs for detection separation times smaller than $|\Delta t_{d}| \lesssim 10^{-21}\;{\rm s}$ with the rest of the distribution uniform. But most disturbingly, the 2D contour plot of the distribution of the event pairs with $R$ and $d$ is now very close to uniform. Clearly, the information in the signal related to the size of the neutrinosphere is all but gone. 

Thus from our three calculations for $|\Delta t_{d}|<10^{-22}$ s, $|\Delta t_{d}|<10^{-21}$ s and $|\Delta t_{d}|<10^{-20}$ s shown in Figs. \ref{fig:MCMC_Histograms_td-22_1D} to \ref{fig:MCMC_Histograms_td-20_2D} we conclude that, for tight enough detection time bounds, the interference signal is clearly present in the two-event spacial distribution. However, as the detection time window is broadened, the interference signal is greatly diminished and in order to extract the information about the size of the source from the signal, very large numbers of event pairs are required. 


\begin{figure}[b]
	\includegraphics[width=0.5\linewidth]{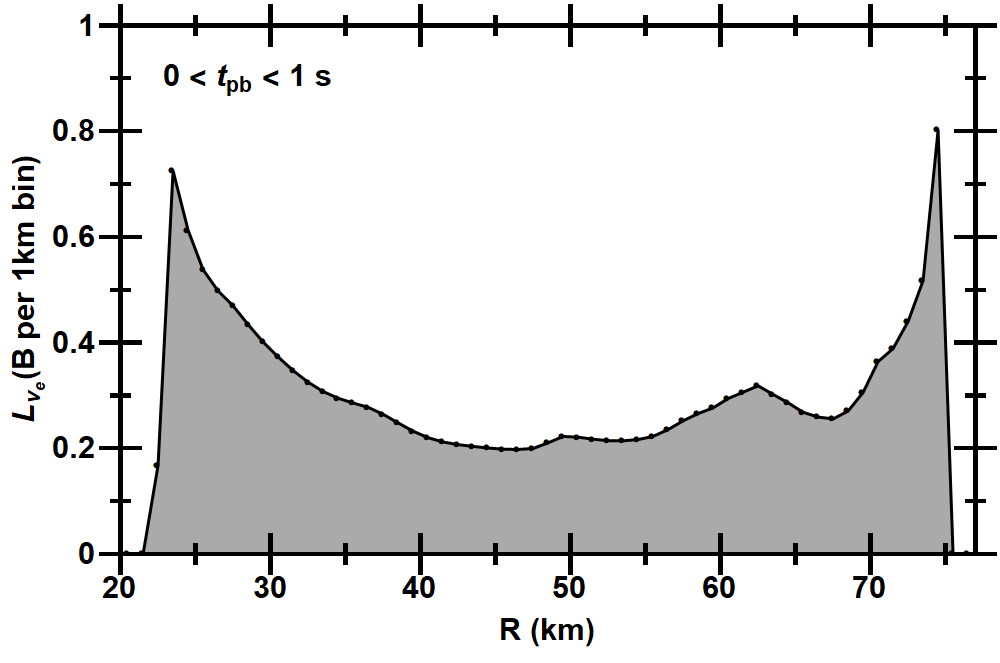}
	\caption{Luminosity as a function of neutrinosphere radius over the first second post bounce.}
	\label{fig:LvsR}
\end{figure}

\begin{figure}[b]
	\includegraphics[width=0.5\linewidth]{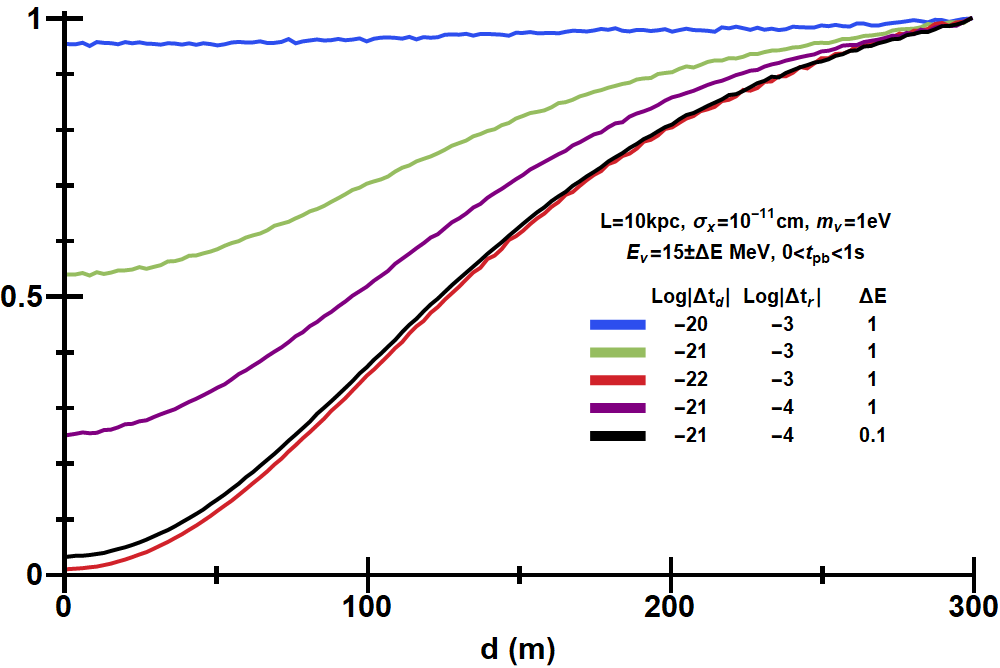}
	\caption{Distribution of events versus event separation.}
	\label{fig:MCMC_Analysis}
\end{figure}

\subsection{Weighted Event Distributions}
One assumption of the previous calculation is clearly flawed: the uniform distribution for the neutrinosphere radius $R$. So let us consider the case where the distribution for the neutrinosphere radius is weighted using a function proportional to the neutrino luminosity versus neutrinosphere radius from a the first second of a hydrodynamical simulation of a core-collapse supernova. The evolution of the radius and luminosity we adopt is taken from the SFHo-z9.6co-nu-1D simulation in \cite{2016NCimR..39....1M}. The relation between these two quantities is shown in Fig. \ref{fig:LvsR}. Using this relation to weight the neutrinosphere radius distribution means our sample is now weighted towards those values of the radius variable $R$ corresponding to those times when the proto-neutron star is emitting lots of neutrinos and thus most likely to produce pairs of events. We now repeat our analysis leaving the distributions for the angles $\theta_i$ and $\phi_i$ as uniform and again considering various neutrino energy, emission and detection time windows. 

The results are shown in Fig. \ref{fig:MCMC_Analysis}. Even with the weighted distribution for $R$, we again find the distribution of events with separation $d$ becomes more uniform as we increase the detection time window, the emission time window, or the neutrino energy window with degeneracies among the three. With an emission time window $|t_{ri}| \leq 10^{-3}\;{\rm s}$ and an energy window of $\Delta E = 1\;{\rm MeV}$, which is the red curve in the figure, the distribution of events has a clear minimum at $d=0$ for the detection time window of $\Delta t_d = 10^{-22}\;{\rm s}$ as previously noted. As we allow for larger detection time windows, the green and the blue curves, but hold the emission time and energy windows fixed, the minimum in the number of events with separations of $d=0$ becomes less deep and by $\Delta t_d = 10^{-20}\;{\rm s}$ it has essentially disappeared. 

As we argued earlier, one would normally expect the emission time windows to be of order the light crossing time of the neutrinosphere which, as Fig. \ref{fig:LvsR} shows, are of order $10^{-4}-10^{-3}\;{\rm s}$. However during the accretion phase of a CCSN significant neutrino emission occurs from hotspots created at the base of downflows onto the proto-neutron star \cite{2016ApJ...818..123B}. Such hotspots would be much smaller than the size of the neutrinosphere so perhaps we can consider smaller time windows. The purple curve in Fig.  \ref{fig:MCMC_Analysis} indicates that if we reduce the time emission window to $10^{-4}\;{\rm s}$ we can compensate for the larger detection time window, and the black curve indicates we can do the same with the energy window. But even with the smaller emission time window and the smaller energy difference, the figure shows the detection time windows needed to see the minimum at $d=0$ are very small.

Thus we conclude that while, in principle, neutrino intensity interferometry could be used to measure the radius of the neutrinosphere, in practice it requires detection time windows which are unfeasibly small. 


\section{Analytic Explanation}
\label{sec:analytical}

The analysis of the neutrino event pair sample we created using the numerical algorithm revealed that in order to observe the spatial variation of the two particle correlation, the time difference between the detected event pairs must be very small. This requirement of an extremely small time difference can be explained analytically. In order to facilitate our explanation we adopt the assumptions of ballistic momenta $(\hat{p}_{ij}=\hat{x}_{ij}=(\vec{d}_j-\vec{r}_i)/\vert(\vec{d}_j-\vec{r}_i)\vert)$ and equal energy $\left(E=E_1=E_2\right)$ and, to further simplify the analysis, we restrict the geometry to a `two-dimensional' case where the emission points $\left(t,x,y,z\right) = \left(0,\pm R,0,0\right)$ and detected locations $\left(L/v,\pm d/2,0,L\right)$ lie in the same spatial plane. 
We define the components of the single-particle wave packets that enter equation (\ref{Eqn:TwoParticleWF}) to be $\psi=N e^{\chi}$ with $\chi$ containing all the important time dependence. Given the symmetry of our setup, the individual $\chi$'s are related and by specifying one, they all can be identified. We choose to define $\chi_{22}$ and the others may be obtained by suitable substitutions:
\begin{align}
\text{Re}\left[\chi_{22}\right] = &\frac{ \gamma^2 \,\left(\gamma ^2-1\right)^2 \, E_\nu^2\,\sigma_x^2\, \left( \xi^{-1} \sqrt{L^2+(d/2-R)^2} -\left( L/\xi-\left(t_{r2}-\Delta t_{d}\right) \right) \right)^2}
{ \left(L/\xi- \left(t_{r2}-\Delta t_{d}\right)\right)^2 +4 \gamma ^4 E_\nu^2 \sigma_x^4 }  \\
\text{Im}\left[\chi_{22}\right]= &  E_\nu\left(
        	\xi\sqrt{L^2+(d/2-R)^2}
            -\left(L/\xi-\left(t_{r2}-\Delta t_{d}\right)\right)
        \right) \nonumber\\
    	& \quad + \frac{E_\nu
        	\left(\gamma ^2-1\right)
        	\left(L/\xi-\left(t_{r2}-\Delta t_{d}\right)\right)
        	\left(
        		\xi^{-1} \sqrt{L^2+(d/2-R)^2}
        		-\left(L/\xi-\left(t_{r2}-\Delta t_{d}\right)\right)
        	\right)^2
        }{
        	2\left(
            	\left(L/\xi-\left(t_{r2}-\Delta t_{d}\right)\right)^2
        		+4 \gamma ^4 E_\nu^2 \sigma_x^4
            \right)
        }.
\end{align}
where $\xi^2=\left(\gamma ^2-1\right)/\gamma^2$.
To obtain $\chi_{11}$ from $\chi_{22}$ substitute $t_{r2}\to t_{r1}$, $\Delta t_{d}\to 0$; for $\chi_{12}$ make the substituions $t_{r2}\to t_{r1}$ and $R\to -R$, and for $\chi_{21}$ replace $R\to -R$ and $\Delta t_{d}\to 0$.
Using these quantities and reasonably asserting that in the astrophysical limit the normalization factors $N$ are non-zero and cancel, we find the 2PCF is given by
\begin{align}
	C_2&=\frac{
    	e^{\Sigma}\left(\cosh\Delta-\cos\theta\right)
    }{
    	e^{\Sigma}\cosh\Delta
    }
\end{align}
where
\begin{align}
\quad\Sigma&=\text{Re}\left[ \chi_{11}
    	+\chi_{12}
        +\chi_{21}
        +\chi_{22}\right],\\
    \Delta&=\text{Re}\left[\chi_{11}
    	-\chi_{12}
        -\chi_{21}
        +\chi_{22}\right],\\
    \theta&=\text{Im}\left[ \chi_{11}
    	-\chi_{12}
        -\chi_{21}
        +\chi_{22} \right].
\end{align}
While the reader will observe the factor $e^{\Sigma}$ occurs in both the numerator and the denominator of the expression for the 2PCF and therefore algebraically cancels. This factor can become extremely small ($\Sigma$ can be very negative) and thus needs careful consideration otherwise we would end up fruitlessly exploring the 2PCF in regions of parameter space where no two-particle events can occur (with or without interference). The final simplifying approximation we make is to assert that all $\text{Re}\left[\chi_{ij}\right]$'s have the same denominator by setting $t_{r1},t_{r2},\Delta t_{d}\to 0$ in the denominators only. We have verified this approximation is valid in the region of parameter space we are considering. In conjunction with the usual assumption of $\gamma^2-1\to\gamma^2$ and by termwise applying the astrophysical limit $\left(L\gg R\gg d\right)$, we find that 
\begin{align}
	\Sigma&=
	\eta_1\,\left(
    	\Delta t_{d}\,(t_{r1}+t_{r2}-\Delta t_{d})-t_{r1}^2-t_{r2}^2
    	-\frac{R^2}{L}\left(t_{r1}+t_{r2}-\Delta t_{d}\right)
    	-\frac{R^4}{2L^2}
    \right),\\
    \Delta&=
	\eta_1\,\left(
    	\Delta t_{d}\,\left(t_{r2}-t_{r1}\right)
    	-\frac{d R}{L\gamma}\left(t_{r1}+t_{r2}-\Delta t_{d}\right)
    	-\frac{d R^3}{L^2}
    \right),\\
    \eta_1 &=\frac{2\gamma^4E_\nu^2\sigma_x^2}{L^2+4\gamma^4E_\nu^2\sigma_x^4} =\frac{1}{2\,\sigma_{\parallel}^2}
\end{align}
For a neutrino with an energy of 15 MeV, $m_\nu=1$ eV, $\sigma_x=10^{-11}$ cm and $L=10$ kpc, the longitudinal spread of the wavefunction of the single-particle wave packet at the detector is $\sigma_{\parallel} \sim 90\;{\rm km}$ as shown in Fig. \ref{fig:WPgrowthParallal}. Since the leading order terms in $\Sigma$ and $\Delta$ are the those quadratic in time - the other terms are very small because they are suppressed by the ratio $R/L$ - we conclude that in order to observe any event pairs at all, the value of the term $\Delta t_{d}\,(t_{r1}+t_{r2}-\Delta t_{d})-t_{r1}^2-t_{r2}^2$ must be greater than $\sim -\sigma^2_{\parallel}/c^2 \sim - 10^{-3}\;{\rm s^2}$ in order to give a value for $e^{\Sigma} \sim 1$. This is not a severe constraint for this set of neutrino mass, energy, initial wave packet size and supernova distance but note that as the neutrino mass decreases, the longitudinal spread of the single-particle wave packet also decreases making the restriction on the combination of emission times and the detection time window more stringent. Similarly, higher neutrino energies, greater initial wave packet size, or smaller distance to the supernova also reduce the size of the wave packet at Earth which also means the absolute value of $\Delta t_{d}\,(t_{r1}+t_{r2}-\Delta t_{d})-t_{r1}^2-t_{r2}^2$ must satisfy a stricter bound. Finally, we note that if value of the terms quadratic in time in $\Sigma$ satisfy this constraint, the term quadratic in time in $\Delta$ is very close to zero. This means that $\cosh\Delta$ is very close to unity and will be assumed to be so for the remainder of this analyses. 

A similar analysis can be performed for $\theta$ and, after making the same simplifying assumptions used for $\Sigma$ and $\Delta$, we find:
\begin{align}\begin{aligned}
	\theta=&
	\eta_2 \bigg(
    	-\frac{3}{2}\Delta t_{d}\, (t_{r1}-t_{r2})\, (t_{r1}+t_{r2}-\Delta t_{d})
    	-\frac{d R}{L}\left(t_{r1}^2+t_{r2}^2+\Delta t_{d}^2\right)
        \\
        &+L\,\Delta t_{d}\,(t_{r1}-t_{r2})
    	-d\, R\,(t_{r1}+t_{r2}-\Delta t_{d})
    \bigg)    +\theta_\text{NII},\\
    \theta_\text{NII}=&\theta_\text{HBT}
    	\left(1+\frac{\gamma ^2 R^2/2}{4\, \gamma^4\, E^2\, \sigma_x^4+L^2}\right) = \theta_\text{HBT}\left( 1 + \frac{ \gamma^2\,R^2}{2\,L^2}\left(\frac{\sigma_{\parallel}^2 - \sigma_{x}^2}{\sigma_{\parallel}^2} \right) \right),\\
    \theta_\text{HBT}=&-2\frac{d E_\nu R}{L} ,\\
    \eta_2=&\frac{\gamma^2\,E_\nu}{L^2+4\,\gamma^4\,E_\nu^2\,\sigma_x^4}=\frac{\gamma^2\,E_\nu}{L^2}\,\left(\frac{\sigma_{\parallel}^2 - \sigma_{x}^2}{\sigma_{\parallel}^2}\right).
\end{aligned}\end{align}
Once again, for 15 MeV energies, $m_\nu=1$ eV, $\sigma_x=10^{-11}$ cm and $L=10$ kpc, the prefactor $\eta_2$ is found to be $\eta_2\sim 10^{12}\;/{\rm s^3}$. Given the astrophysical limit and the expectation of the emission times being of order $\sim10^{-3}$ s, it is the term in $\theta$ which is linear in $L$ which contributes the most to the difference between $\theta$ and $\theta_\text{NII}$. As the detection time window increases the possible values of the product $\eta_2\,L\,\Delta t_d (t_{r1}-t_{r2})$ also become larger leading to ever larger shifts of $\theta$ away from $\theta_{NII}$. The greater the possible shifts from $\theta_{NII}$ the wider the spread of the $\cos\theta$ term which appears in the 2PCF. In order to observe the interference, $\theta$ cannot differ greatly from $\theta_{NII}$ which cannot differ greatly\footnote{The requirement that $\theta_{NII}$ not differ greatly from $\theta_{HBT}$ was a constraint established by W\&K.} from $\theta_{HBT}$. 

\begin{figure}[ht]
	\includegraphics[width=0.45\linewidth]{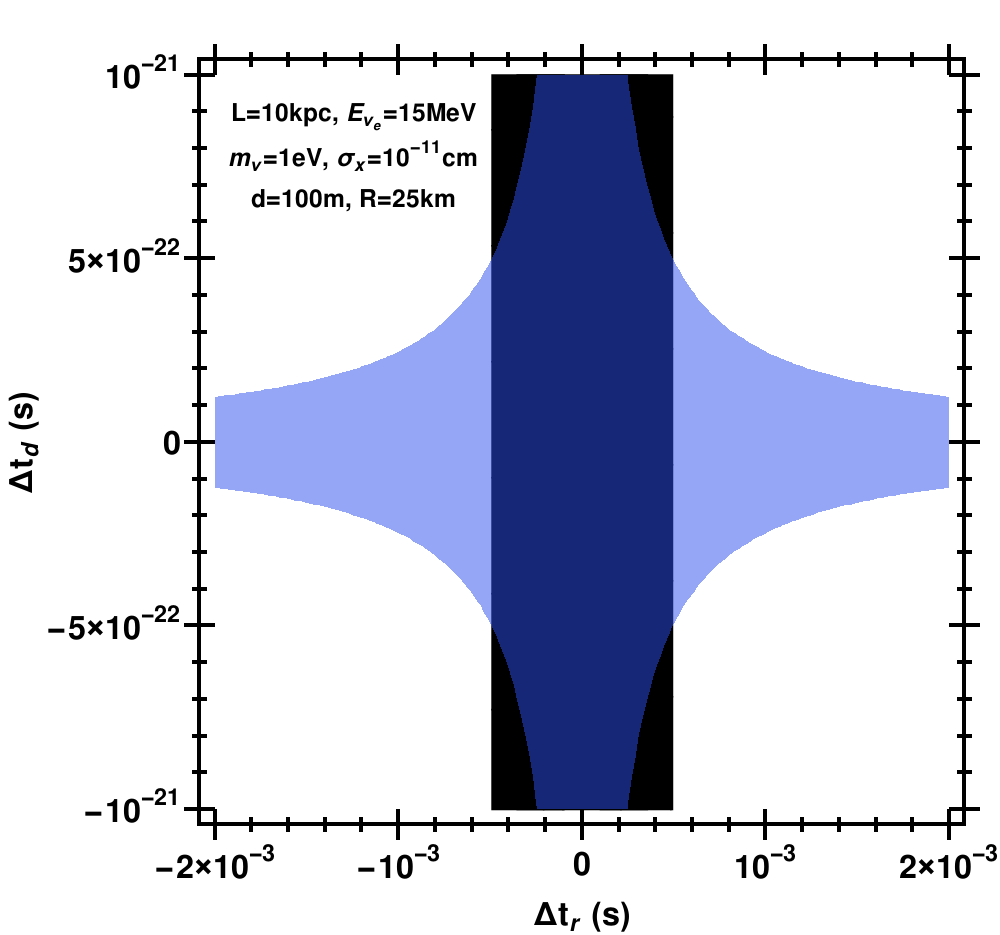}
	\includegraphics[width=0.45\linewidth]{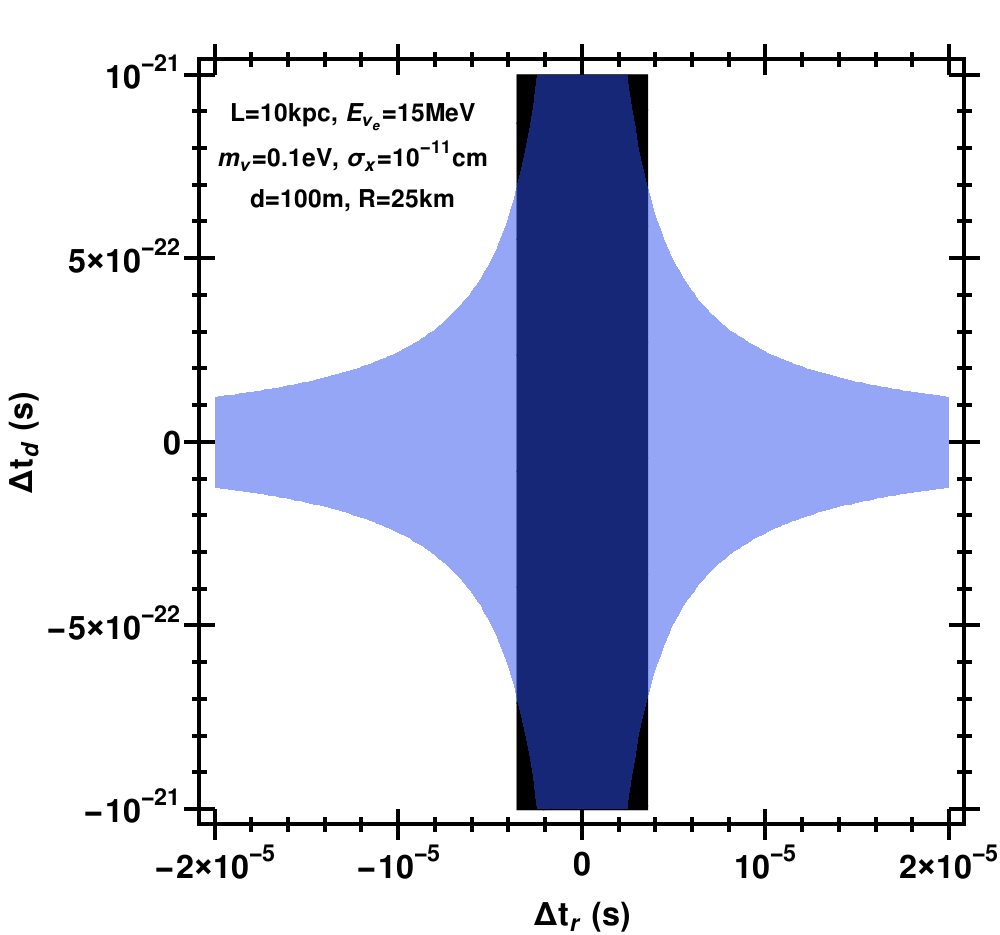}
	\caption{Allowed regions defined by Eq. \ref{Eqn:AnalyticContraints}.}
	\label{fig:TimeAnalysis}
\end{figure}

The above analysis gives rise to the following two conditions that together ensure that the 2PCF is a useful signal which we define to be a reasonable possibility of producing a correlated pair of neutrino events, and values for $\theta$ which do not differ greatly from $\theta_{HBT}$:
\begin{align}\begin{aligned}
	\text{Exp}\left[\eta_1 \,\left(\Delta t_{d}\, (t_{r1}+t_{r2}-\Delta t_{d})-t_{r1}^2-t_{r2}^2\right) \right]&>\frac{1}{2}\\
    \frac{\eta_2\, L\,\Delta t_{d}\, (t_{r1}-t_{r2})}{\theta_\text{HBT}}&<1
    \label{Eqn:AnalyticContraints}
\end{aligned}\end{align}
The first inequality ensures that at least in half the cases of two neutrino emission, the single-particle wave packets of the neutrinos are overlapping in the detector and could give rise to a correlated pair. The second inequality is the requirement that the interference pattern is not washed out by large variations of the $\theta$ term. The first constraint is very aggressive - one can tolerate fewer overlapping wave packets i.e. a smaller right-hand-side of the inequality - if the source were static (such as a star) because the decreased probability of event pairs can be compensated by longer exposure times. However this is not possible for a transient source such as a supernova. Even if we relax this requirement, the constraint in the time window is only logarithmic in the probability of having the overlapping wave packets. The two constraints are shown in both plots of Fig. \ref{fig:TimeAnalysis}. The left plot is for $m_\nu=1$ eV and the right plot is for $m_\nu=0.1$ eV. The black (blue) region is the allowed region defines by the first (second) inequality in Eq. \ref{Eqn:AnalyticContraints}. The limits of $\Delta t_r=t_{r1}-t_{r2}$ are chosen to that the normalization constraints are visible (from the first inequality in Eq. \ref{Eqn:AnalyticContraints}). The limits on $\Delta t_{d}$ are selected such that the blue region almost entirely covers the black region i.e. the region where the interferometric signal is useful overlaps the region where two-particle events can occur. These bounds remain qualitatively unchanged when event separation and source size are varied over appropriate intervals (given in Sec. \ref{sec:MCMC_Description}). The difference between the two plots gives an indication of the effect of neutrino mass. When mass is decreased by a factor of 10, the bounds on source emission time for overlapping wave packets is decreased by a factor of $\sim$100. While this doesn't effect the detector time resolution requirements by much, it does mean that getting correlated pairs is 100 times harder. Thus we find that in order to satisfy the inequalities in Eq. \ref{Eqn:AnalyticContraints}, the experimental time resolution required for useful interferometric signal is extreme, $\Delta t_d \lesssim 10^{-21}\;{\rm s}$ for $m_{\nu} = 1\;{\rm eV}$, $E=15\;{\rm MeV}$, $L=10\;{\rm kpc}$ and $\sigma_x = 10^{-11}\;{\rm cm}$ which is the same as we found numerical analysis. 
Figure \ref{fig:TimeAnalysis} shows that as the neutrino mass decreases, a larger window of $\Delta t_{d}$ is allowed but the window on $\Delta t_r$ shrinks. This behavior can be explained from the two inequalities. The second inequality from Eq. \ref{Eqn:AnalyticContraints} can be re-arranged using the definitions of $\theta_{HBT}$, $\eta_2$ and $\sigma_{\parallel}$ to give
\begin{align}\begin{aligned}
	\Delta t_{d} & < \frac{2\,d\,R}{\gamma^2\,(t_{r1}-t_{r2})}\,\left(1+ \frac{4\gamma^4\,\sigma_x^4 E_{\nu}^2}{L^2}\right)
\end{aligned}\end{align}
When the term $4\gamma^4\,\sigma_x^4 \,E_{\nu}^2  / L^2 \ll 1$, i.e.\hspace{-2mm} larger neutrino masses, the detector timing constraint is $\Delta t_{d}  < (2\,d\,R )/ \gamma^2\,(t_{r1}-t_{r2})$. In this limit $\Delta t_{d} \ll t_{ri}$ so the difference between the emission times, $(t_{r1}-t_{r2})$, is of order $\sigma_{\parallel}/c$. Thus $\Delta t_{d}$ is set by the ratio $ d R / c \sigma_{\parallel}$ divided by the square of the Lorentz factor of the neutrino - which is very large if the neutrino mass $m_{\nu}$ is of order 1 eV or less and the neutrino energy is $E_{\nu} = 15\;{\rm MeV}$. When the term $4\gamma^4\,\sigma_x^4\, E_{\nu}^2  / L^2 \gg 1$, which occurs as the neutrino mass approaches zero, we find $\Delta t_{d}  < (8\,d\,R\,\,\gamma^2\,\sigma_{x}^4\,E_{\nu}^2)/(t_{r1}-t_{r2}) L^2$. Now the time detection window constraint from the second inequality is expanded by the Lorentz factor. In the limit $m_{\nu} \to 0$, the detection time window constraint from the second inequality is always satisfied but, as $m_{\nu} \to 0$, it becomes harder to satisfy the constraint from the first inequality on the number of correlated pairs because the longitudinal spread of the wave packet at Earth decreases. For massless neutrinos, $\sigma_{\parallel} = \sigma_x$ at Earth so the combination $\Delta t_{d}\, (t_{r1}+t_{r2}-\Delta t_{d})-t_{r1}^2-t_{r2}^2$ needs to be of order the square of the light-travel time of the initial neutrino wave packet i.e. $\sigma_x^2/c^2$. For $\sigma_{x} = 10^{-11}\;{\rm cm}$ this time is $\sigma_{x} /c = 3\times 10^{-22}\;{\rm s}$. In the massless neutrino limit, the number $N_{2\nu}$ of overlapping wavepackets is $N_{2\nu} \sim 10^{-7}\;{\rm /m^2}$ for a supernova at $L=10\;{\rm kpc}$.

Thus we find the two requirements in Eq. \ref{Eqn:AnalyticContraints} are `orthogonal' in the sense that changing parameters so that it becomes easier to satisfy one makes it more difficult to satisfy the other. There is really no way to evade the bound that the detection time window has to be extremely small given the initial wave packet size, realistic distances to supernovae, neutrino emission over a period of seconds, and detector dimensions which are measured in meters. 

\section{Conclusion \label{sec:Conclusion}}

In this paper we have continued the study of neutrino intensity interferometry by relaxing the assumptions that were used in Wright \& Kneller. These were: the requirement of equal times of emission and detection, the assumption of equal energies for the two neutrinos, and the assumption that the two points of emission and two points of detection lie in plane. While the relaxation of all these assumptions generally reduces the significance of the correlation signal, it is the relaxation of the assumption of equal times of detection that leads to the greatest loss and the principle reason why neutrino intensity interferometry becomes difficult to realize in practical terms. Unfortunately our analysis indicates this conclusion is robust. For neutrinos with an initial wave packet spread of $\sigma_x \sim 10^{-11}\;{\rm cm}$ and energies of $E \sim 15\;{\rm MeV}$ emitted from a supernova at $L=10\;{\rm kpc}$, a neutrino mass greater than $m_{\nu} \sim 10^{-9}\;{\rm eV}$ means the detection time window must be smaller than $d R / ( \gamma^2 \sigma_{\parallel} )$ where $d$ is the distance between the detected pair, $R$ is the radius of the source, $\sigma_{\parallel}$ is the longitudinal spread of the neutrino wavepacket at Earth, and $\gamma$ is the neutrino Lorentz factor. For smaller neutrino masses the detection time window must be smaller than the light-travel time of the initial neutrino wave packet. Both times are of order $\sim 10^{-21}\;{\rm s}$ or less.  

Finally, the reader may be curious why intensity interferometry works for photon pairs when measuring the sizes of stars and not for neutrinos from supernovae. The key difference is the size of the photon wave packet. For a typical main sequence star the mean free path of an atom in its atmosphere is of order $\sim10^{-4}\text{ m}$ which, together with a typical thermal velocity, gives a time between collisions of $\sim10^{-8}\text{ s}$. This means any photon produced by emission in that environment has a wave packet size of $\sigma_x\sim10^2\text{ cm}$. One can also estimate that the coherence time of continuous bremsstrahlung emission via free electrons near $\text{H}^-$ ions in the stellar photosphere and find $\sim10^{-11}\text{ s}$ which would give a wave packet size $\sigma_x\sim0.4\text{ cm}$. Thus both emission processes indicate the initial size of the photon wave packet is much larger than the neutrino's emitted from the neutrinosphere in a core collapse supernova and this greater size makes the technique feasible for a static and sufficiently bright source if one uses detectors with a wavelength resolution of $\Delta \lambda \lesssim 30\;{\rm nm}$. 


\section*{Acknowledgments}
\noindent 
This work was supported at NC State by DOE grants DE-FG02-02ER41216 and SC0010263.

\appendix 
\section{MCMC Coverage and Convergence \label{appendix:MCMC_CandC}}
\begin{figure}[ht]
	\includegraphics[width=0.5\linewidth]{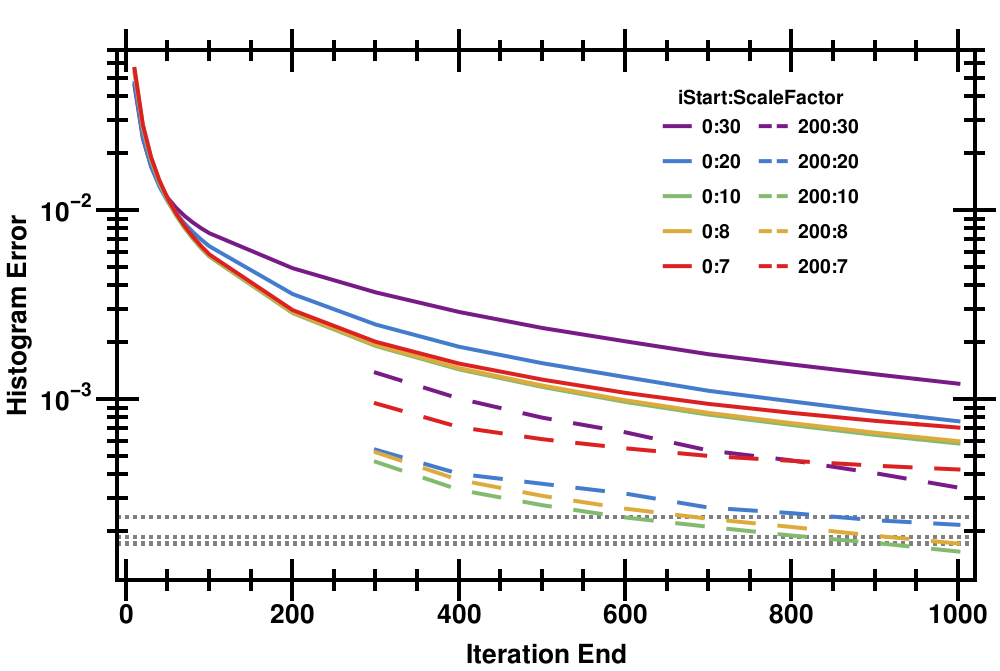}
	\caption{Distribution of events versus event separation.}
	\label{fig:MCMC_Convergence}
\end{figure}

In order to analyze the convergence of the MCMC and the its sensitivity to the scale factor, we consider the 2D histogram of $R$ and $d$ and examine its behavior as a function of both iteration and scale factor. The results are presented in Fig. \ref{fig:MCMC_Convergence}. This figure has the iteration number on the x-axis and a quantification of the histogram error on the y-axis. The histogram error is a measure of how far the 2D histogram deviates from an ensemble of converged histograms. The various curves displayed in Fig. \ref{fig:MCMC_Convergence} are labeled in the legend. The label "iStart" indicates from which iteration the histogram accumulates data (for a particular point on the line, the corresponding point on the x-axis indicates the iteration at which the accumulation ends). The label "Scale Factor" denoted what scale factor is used for each line (see Sec. \ref{sec:MCMC_Description}). Figure \ref{fig:MCMC_Convergence} also indicates, as dashed lines, the level of error the ensemble of converged histograms have with each other as a quantification of when a MCMC can be considered as converged. Thus the figure shows quite clearly that by burning the first 200 iterations, the time-to-convergence is drastically reduced. Furthermore, we see that the red curve, with scale factor of 30, and the purple curve, with scale factor of 7, have a slower convergence rate as compared to the other three curves with scale factors between 7 and 30. Most of our numerical calculations were performed with a burn of 250 iterations and a scale factor of 20. Figure \ref{fig:MCMC_Convergence} shows that such a choice will converge well before the 1000 iterations all of our calculations were performed for. This gives confidence that our choice of burn count and scale factor yield stable results and that the final, analyzed histograms are in a converged state.

\bibliographystyle{apsrev4-1}
\bibliography{main}

\end{document}